\documentclass[aps,prb,reprint,preprintnumbers,superscriptaddress,amsmath,amssymb,bibnotes,longbibliography]{revtex4-1}

\usepackage{graphicx}% Include figure files
\usepackage{dcolumn}% Align table columns on decimal point
\usepackage{bm}% bold math
\usepackage[colorlinks,linkcolor=blue,anchorcolor=blue,citecolor=blue,urlcolor=blue,filecolor=blue,menucolor=blue,runcolor=blue]{hyperref}
\usepackage{ulem}% add hypertext capabilities
\begin{document}

\title{Full superconducting gap and type-I to type-II superconductivity transition in single crystalline NbGe$_2$}% Force line breaks with \\

\author{Dongting Zhang}
\author{Tian Le}
\affiliation  {Center for Correlated Matter and Department of Physics, Zhejiang University, Hangzhou 310058, China}
\author{Baijiang Lv}
\affiliation  {Department of Physics, Zhejiang University, Hangzhou 310027, China}
\affiliation  {Zhejiang Province Key Laboratory of Quantum Technology and Device, Zhejiang University, Hangzhou 310027, China}
\author{Lichang Yin}
\author{Chufan Chen}
\affiliation  {Center for Correlated Matter and Department of Physics, Zhejiang University, Hangzhou 310058, China}
\author{Zhiyong Nie}
\author{Dajun Su}
\affiliation  {Center for Correlated Matter and Department of Physics, Zhejiang University, Hangzhou 310058, China}
\author{Huiqiu Yuan}
\affiliation  {Center for Correlated Matter and Department of Physics, Zhejiang University, Hangzhou 310058, China}
\affiliation  {Zhejiang Province Key Laboratory of Quantum Technology and Device, Zhejiang University, Hangzhou 310027, China}
\affiliation  {State Key Laboratory of Silicon Materials, Zhejiang University, Hangzhou 310027, China}
\affiliation  {Collaborative Innovation Center of Advanced Microstructures, Nanjing University, Nanjing 210093, China}
\author{Zhu-An Xu}
\affiliation  {Department of Physics, Zhejiang University, Hangzhou 310027, China}
\affiliation  {Zhejiang Province Key Laboratory of Quantum Technology and Device, Zhejiang University, Hangzhou 310027, China}
\affiliation  {State Key Laboratory of Silicon Materials, Zhejiang University, Hangzhou 310027, China}
\affiliation  {Collaborative Innovation Center of Advanced Microstructures, Nanjing University, Nanjing 210093, China}
\affiliation  {Zhejiang California International NanoSystems Institute, Zhejiang University, Hangzhou 310058, China}
\author{Xin Lu}
\email[Corresponding author: ]{xinluphy@zju.edu.cn}
\affiliation  {Center for Correlated Matter and Department of Physics, Zhejiang University, Hangzhou 310058, China}
\affiliation  {Zhejiang Province Key Laboratory of Quantum Technology and Device, Zhejiang University, Hangzhou 310027, China}
\affiliation  {Collaborative Innovation Center of Advanced Microstructures, Nanjing University, Nanjing 210093, China}

\date{\today}% It is always \today, today,

\begin{abstract}
We report a mechanical point-contact spectroscopy study on the single crystalline NbGe$_2$ with a superconducting transition temperature $T\rm_c$ = 2.0 - 2.1 K. The differential conductance curves at 0.3 K can be well fitted by a single gap s-wave Blonder-Tinkham-Klapwijk model and the temperature dependent gap follows a standard Bardeen-Cooper-Schrieffer behavior, yielding $\Delta_0 \sim$ 0.32 meV and 2$\Delta_0$/$k\rm_{B}$$T\rm_{c}$ = 3.62 in the weak coupling limit. In magnetic field, the superconducting gap at 0.3 K keeps constant up to $H_{c1}\sim$150 Oe and gradually decreases until $H_{c2}\sim$350 Oe, indicating NbGe$_2$ going through a transition from type-I to type-II (possible type-II/1) superconductor at low temperature.
\end{abstract}

%pacs{Valid PACS appear here}% PACS, the Physics and Astronomy
\maketitle

Noncentrosymmetric materials have attracted intensive attention in recent years, whose absence of inversion symmetry in the crystal unit cell can induce an antisymmetric spin-orbit coupling (ASOC) and nontrivial topology of electronic bands \cite{annurev-conmatphys-031113-133912, 1361-6633/80/3/036501, PhysRevB.101.245145}. If additional mirror or other roto-inversion symmetries are broken in such materials, they form a unique class of topological chiral crystals, and are proposed to host Kramers-Weyl fermions \cite{hlca.200390109, s41563-018-0169-3}. This new type of topological fermions appear at time-reversal-invariant momenta with a broad energy range of nontrivial topological bands, and these fermions are connected by Fermi arcs with a large spanning length in the reciprocal space \cite{s41563-018-0169-3, s41563-018-0210-6, s41586-019-1037-2}. For example, topological bands in the RhSi family have been experimentally observed with characters of chiral crystals, and their helicoid Fermi arcs on surface are claimed to have a Chern number of {$\pm$} 2 \cite{s41586-019-1037-2}. On the other hand, noncentrosymmetric superconductors can in principle allow an admixture of spin-singlet and -triplet parings \cite{427799a, annurev-conmatphys-031113-133912, 1361-6633/80/3/036501}, and serve as a promising platform to realize intrinsic topological superconductors, such as PbTaSe$_2$ \cite{PhysRevB.89.020505, sciadv.1600894, j.scib.2020.04.039} and BiPd \cite{PhysRevB.86.094520, ncomms13315, PhysRevB.99.020507, PhysRevLett.124.167001}. Among them, noncentrosymmetric NbGe$_2$ has been reported to have a superconducting transition temperature $T_c \sim$ 2.09 {$\pm$} 0.02 K \cite{0022-5088(78)90033-4} decades ago. Interests on NbGe$_2$ have reemerged due to a recent proposal as a chiral crystal candidate, where its superconductivity and nontrivial topology may be intricately intertwined \cite{s41563-018-0169-3}.

In addition, recent magnetization and specific heat measurements on NbGe$_2$ have confirmed its crossover from type-I to type-II superconductor with decreased temperatures \cite{PhysRevB.102.064507, PhysRevB.102.235144}, similar to the case of ZrB$_{12}$ and LaRhSi$_3$ \cite{PhysRevB.72.024548, JPSJ.85.024715}. In general, superconductors can be simply classified into type-I and type-II cases, where the Meissner state becomes normal suddenly at H $>$ $H_{c}$ for type-I superconductors but magnetic field can enter into the sample continuously in the form of quantum flux in type-II superconductors with a transition from the Meissner phase to Shubnikov phase. The Ginzburg-Landau parameter $\kappa = \frac{\lambda}{\xi}$ can be a good indicator, where $\kappa$ for type-I superconductor is smaller than $1/{\sqrt{2}}$ and $\kappa$ $>$ $1/{\sqrt{2}}$ for type-II case. However, when the $\kappa$ value is close to $1/{\sqrt{2}}$, an intermediate mixed (I-M) state can exist between the Meissner state and the mixed state (Shubnikov phase) \cite{JPSJ.85.024715, PhysRevLett.102.136408, ncomms9813, PhysRevB.100.064503}. It has been referred as type-II/1 superconductor in order to be distinguished from the conventional type-II superconductor (type-II/2 superconductor) \cite{PhysRevB.7.136}. For type-II/1 superconductor, a discontinuous increase of the flux density from zero to a certain value $B_{0}$ corresponds to a first-order phase transition from the Meissner state to I-M state at $H_{c1}$ \cite{PhysRevB.7.136, PhysRevB.72.024548,BF02763399}. The magnetization usually shows an abrupt decrease from the $4\pi M = -H$ line at $H_{c1}$ with a long tail till $M = 0$ at $H_{c2}$, suggesting the I-M and mixed states, respectively \cite{PhysRevB.7.136, PhysRevB.72.024548, PhysRevB.102.064507}. It is thus desirable to systematically explore the possible topological superconductivity in NbGe$_2$ and its superconducting nature.

In this article, we have applied mechanical point-contact spectroscopy (MPCS) to investigate the superconducting gap in single crystalline NbGe$_2$. The conductance curves for MPCS on NbGe$_2$ at 0.3 K can be well fitted by the Blonder-Tinkham-Klapwijk (BTK) model with a single s-wave gap, where the superconducting gap $\Delta$ follows a typical Bardeen-Cooper-Schrieffer (BCS) temperature behavior, yielding $\Delta_0 \sim$ 0.32 meV and 2$\Delta_0$/$k\rm_{B}$$T\rm_{c}$ = 3.62. The Andreev reflection signal in conductance curves keeps the same in magnetic field up to 150 Oe, and is gradually suppressed until its upper critical field 350 Oe, consistent with a type-II superconductor at low temperatures.

\begin{figure}
\includegraphics[angle=0,width=0.49\textwidth]{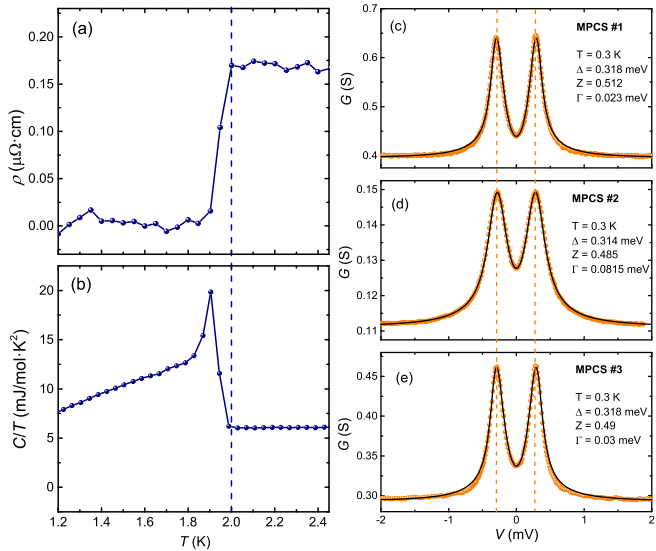}
\vspace{-12pt} \caption{\label{Figure1}(color online) (a) and (b) Temperature-dependent electrical resistivity $\rho$, and specific heat C/T of NbGe$_2$, respectively; (c)-(e) A representative set of point-contact conductance curves on NbGe$_2$ at 0.3 K from MPCS, in comparison with a single gap s-wave BTK fitting (black lines).}
\vspace{-12pt}
\end{figure}

NbGe$_2$ single crystals were grown by a two-step vapor transport technique, where iodine is the transport agent with stoichiometric amounts of high-purity niobium (99.99\%) and germanium (99.99\%) as described elsewhere \cite{PhysRevB.102.064507}. The electrical resistivity and specific-heat of NbGe$_2$ were measured by a Physical Property Measurement System (PPMS) from Quantum Design with a Helium 3 insert cooling down to 0.4 K. Both the specific-heat jump and resistive drop at $T\rm_{c}$ $\sim$ 2 K support the high quality of the NbGe$_2$ crystals, as shown in Fig. \ref{Figure1}(a) and (b). The sharp specific jump in Fig. \ref{Figure1}(b) is probably due to a remnant field in PPMS causing a first-order transition, consistent with the reported specific-heat behavior in field \cite{PhysRevB.102.064507, PhysRevB.102.235144}. Mechanical PCS in a needle-anvil type was employed to study the superconducting gap of NbGe$_2$, where a sharp gold tip prepared by the electrochemical etching was gently engaged on the crystal surface by piezoelectric-controlled nanopositioners. For the point-contact, its differential conductance curves as a function of bias voltage, G(V), were measured in a quasi-four-probe configuration by the conventional lock-in technique, where the experimental details can be referred to \onlinecite{Tortello_2016, 9372}. The ac susceptibility of NbGe$_2$ was measured in coils by generating an ac field of 1337 Hz in frequency and 0.8 Oe in amplitude, and both the dc and ac fields are perpendicular to the polished sample surface along the $\it{c}$ axis. Oxford cryostat with a He 3 insert was used for MPCS and ac susceptibility measurements to cool the sample down to 0.3 K and to apply magnetic field up to 1000 Oe.

\begin{figure}
\includegraphics[angle=0,width=0.49\textwidth]{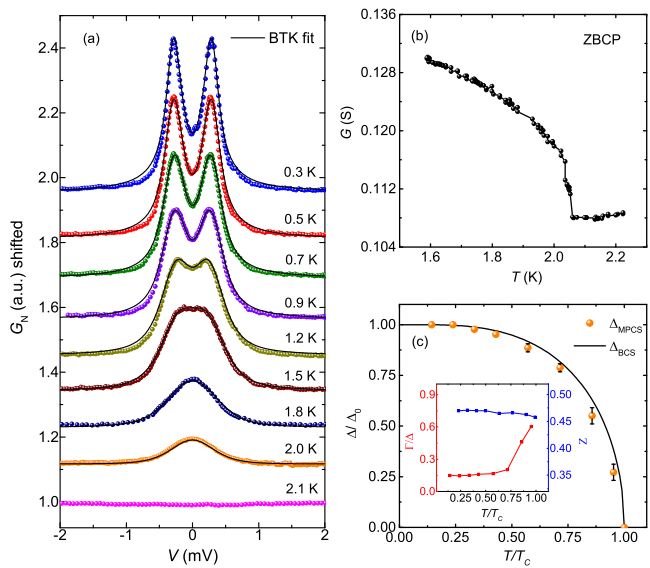}
\vspace{-12pt} \caption{\label{Figure2}(color online) (a) Temperature evolution of the normalized differential conductance curves $G\rm_{N}$(V) from 0.3 to 2.1 K for MPCS on NbGe$_2$, in comparison with a single gap s-wave BTK fitting (black lines). The curves are shifted vertically for clarity. (b) The MPCS zero-bias conductance as a function of temperature with a kink at $T_c\sim$ 2 - 2.1 K. (c) Temperature dependence of the extracted superconducting gap $\Delta$ from the single-gap BTK fitting (solid circle) in comparison with the standard BCS temperature curve (black line). The inset shows the fitting parameters Z and $\Gamma$/$\Delta$ as a function of temperature.}
\vspace{-12pt}
\end{figure}

Figure \ref{Figure1}(c)-(e) show a representative set of differential conductance curves G(V) at 0.3 K for three different contacts on NbGe$_2$. All the conductance curves have a common double-peak structure and can be well fitted by a single-gap s-wave BTK model, suggesting a full superconducting gap in NbGe$_2$. The obtained gap $\Delta$ $\sim$ 0.316 meV at 0.3 K and the smearing factor $\Gamma$ varies between 0.02 and 0.08 meV, signaling a small scattering rate and thus a clean interface for our point-contacts. A slight deviation of the experimental conductance curve from the BTK fitting can be noticed for Fig. \ref{Figure1}(e) at high bias voltages but should not undermine our analysis \cite{PhysRevB.69.134507}. The temperature evolution of point-contact conductance curves are shown in Fig. \ref{Figure2}(a) from 0.3 K to 2.1 K, where the double peaks gradually shift to zero-bias with increased temperatures and the G(V) curve becomes flat at 2.1 K for NbGe$_2$ in the normal state. If we track the temperature dependent zero-bias conductance (ZBC) of MPCS on NbGe$_2$ as in Fig. \ref{Figure2}(b), a kink at 2.05 K in the ZBC curve indicates the disappearance of Andreev reflection and that the NbGe$_2$ crystal transforms from superconducting into normal state, consistent with the resistivity and specific heat data. For all the measured contacts on NbGe$_2$, they have a narrow range of superconducting transition temperatures $T\rm_{c}=$ 2.0 - 2.1 K. The extracted superconducting gap values $\Delta$ from the BTK fitting are plotted in Fig. \ref{Figure2}(c) as a function of the reduced temperature $T/T_c$ and they follow the typical BCS temperature behavior, yielding $\Delta_0$ $\sim$ 0.32 meV and 2$\Delta_0$/$k\rm_{B}$$T\rm_{c}$ = 3.62 in the weak-coupling regime, close to the reported value 3.528 from the specific heat \cite{PhysRevB.102.235144}. We note that the tunneling barrier parameter Z keeps almost constant while the ratio between the smearing parameter $\Gamma$ and superconducting gap $\Delta$, $\Gamma$/$\Delta$, sharply increases close to $T\rm_{c}$ as in the inset of Fig. \ref{Figure2}(c) due to the enhanced pair-breaking effect \cite{0953-2048/23/4/043001}.

\begin{figure}
\includegraphics[angle=0,width=0.49\textwidth]{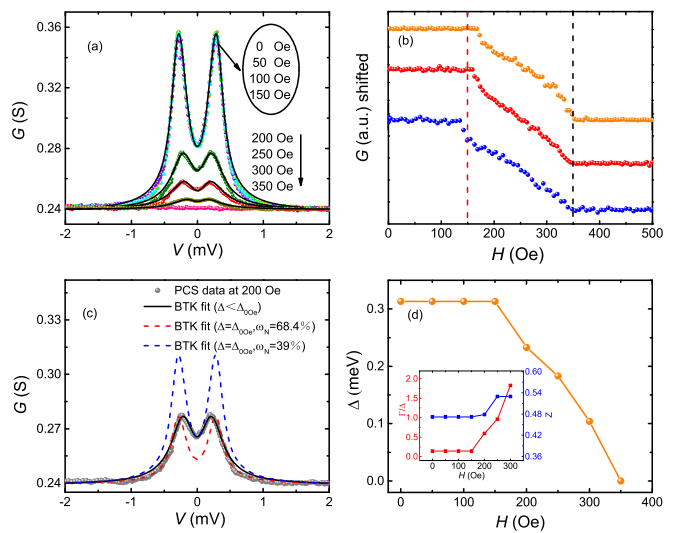}
\vspace{-12pt} \caption{\label{Figure3} (Color online) (a) Field evolution of the differential conductance curves $G(V)$ from zero to 350 Oe for MPCS on NbGe$_2$ at 0.3 K, in comparison with the single gap s-wave BTK fitting curves (black lines). (b) The zero bias conductance of three different contacts on on NbGe$_2$ at 0.3 K as a function of field. The upper critical field is marked by a black dashed line while the red dashed line marks the lower critical field $H\rm_{c1}$. (c) Different Fitting curves with the single-gap or two-component BTK model by varying $\omega$ (68.4\% and 39\%) in comparison with the differential conductance curve at 200 Oe. (d) Field dependence of the superconducting gap $\Delta$ extracted from the single-gap BTK fitting. The inset shows the fitting parameters Z and $\Gamma$/$\Delta$ as a function of field.}
\vspace{-12pt}
\end{figure}

The point-contact conductance curves at 0.3 K shows an interesting behavior in magnetic field as in Fig. \ref{Figure3}(a): They do not change with magnetic field at all up to 150 Oe, characteristic of the Meissner state in field, and the conductance peak intensity is dramatically suppressed by field above 150 Oe, in reminiscence of our previous MPCS observations on the type-I superconductor PdTe$_2$ \cite{PhysRevB.99.180504}. However, in contrast to the case of PdTe$_2$, the peak positions shift to zero-bias voltage and the double-peak distance gradually decreases, similar to the behavior of a conventional type-II superconductor in its mixed state instead \cite{2053-1591/aae2eb, PhysRevB.99.180504}. The Andreev reflection signal is totally suppressed and the curves become flat above its upper critical field $H_{c2}\sim$ 350 Oe. The gradual decrease of double-peak distance supports NbGe$_2$ as a type-II superconductor at low temperature 0.3 K, implying a transition from type-I to type-II superconductivity with decreased temperature as reported for NbGe$_2$ \cite{PhysRevB.102.064507, PhysRevB.102.235144}. The ZBC curves as a function of field for three different contacts are plotted in Fig. \ref{Figure3}(b) with a consistent manner, where their conductance values keep constant up to 150 Oe and decrease until its upper critical field $H_{c2}\sim$ 350 Oe. The ZBC behavior is similar to the results on type-I superconductors Al, Sn (as shown in the Supplemental Material \cite{SI})  and PdTe$_2$ \cite{PhysRevB.99.180504}, however, we note the field range for the conductance change of NbGe$_2$ ($\sim$200 Oe) is much broader in comparison.

In order to argue against the scenario that the smooth decrease of the conductance peak intensity above 150 Oe originates from the intermediate state for a type-I superconductor as in PdTe$_2$ \cite{PhysRevB.99.180504}, we have tried to analyze NbGe$_2$ conductance data at 200 Oe with the same two-component BTK model $G(V)$ = $\omega$$G_{Normal}$ + (1-$\omega$)$G\rm_{SC}$$(V)$, which works well for soft PCS on type-I superconductors PdTe$_2$ \cite{PhysRevB.99.180504} and Al (Refer to FigS1 in the Supplemental Material \cite{SI}). In the intermediate state of a type-I superconductor, there exists a phase separation between the Meissner superconducting state and normal state due to a large demagnetization factor. The weighting parameter $\omega$ for the normal state is proportional to its volume fraction in the contact area and the conductance $G\rm_{SC}$$(V)$ from the superconducting part should gave the same shape as in the zero-field case, where the gap $\Delta$, $\Gamma$ and Z are constant in the Meissner state. In PdTe$_2$, the experimental curves $G(V)$ can be well fitted by only changing the spectra weight $\omega$ for different fields near $H_c$, implying an intermediate state for the type-I superconductor \cite{PhysRevB.99.180504}. However, the same procedure fails when fitting the conductance curve of NbGe$_2$ at 200 Oe, with respect to both the peak intensity and position, as shown in Fig. \ref{Figure3}(c) by the dashed lines. In comparison, the conductance curve at 200 Oe can be well fitted by the BTK model with a reduced gap value, suggesting NbGe$_2$ in the type-II-like mixed state instead. In the pure Meissner state of NbGe$_2$ below 150 Oe, $\omega$ is simply zero and thus the total conductance G(V) would not change. Our mechanical point-contact results in field would thus favor NbGe$_2$ as a type-II rather than type-I superconductor at 0.3 K. In both the Meissner state below 150 Oe and mixed state between 150 and 350 Oe, the conductance curves for point-contact on NbGe$_2$ can be well fitted by the one-gap s-wave BTK model as in Fig. \ref{Figure3}(a) and the extracted parameters are shown in Fig. \ref{Figure3}(d) and its inset, where the superconducting gap keeps constant below 150 Oe and decreases to zero at $H_{c2}$ while the $\Gamma$/$\Delta$ dramatically increases above 150 Oe with flux entering the NbGe$_2$ crystal.

\begin{figure}
\includegraphics[angle=0,width=0.49\textwidth]{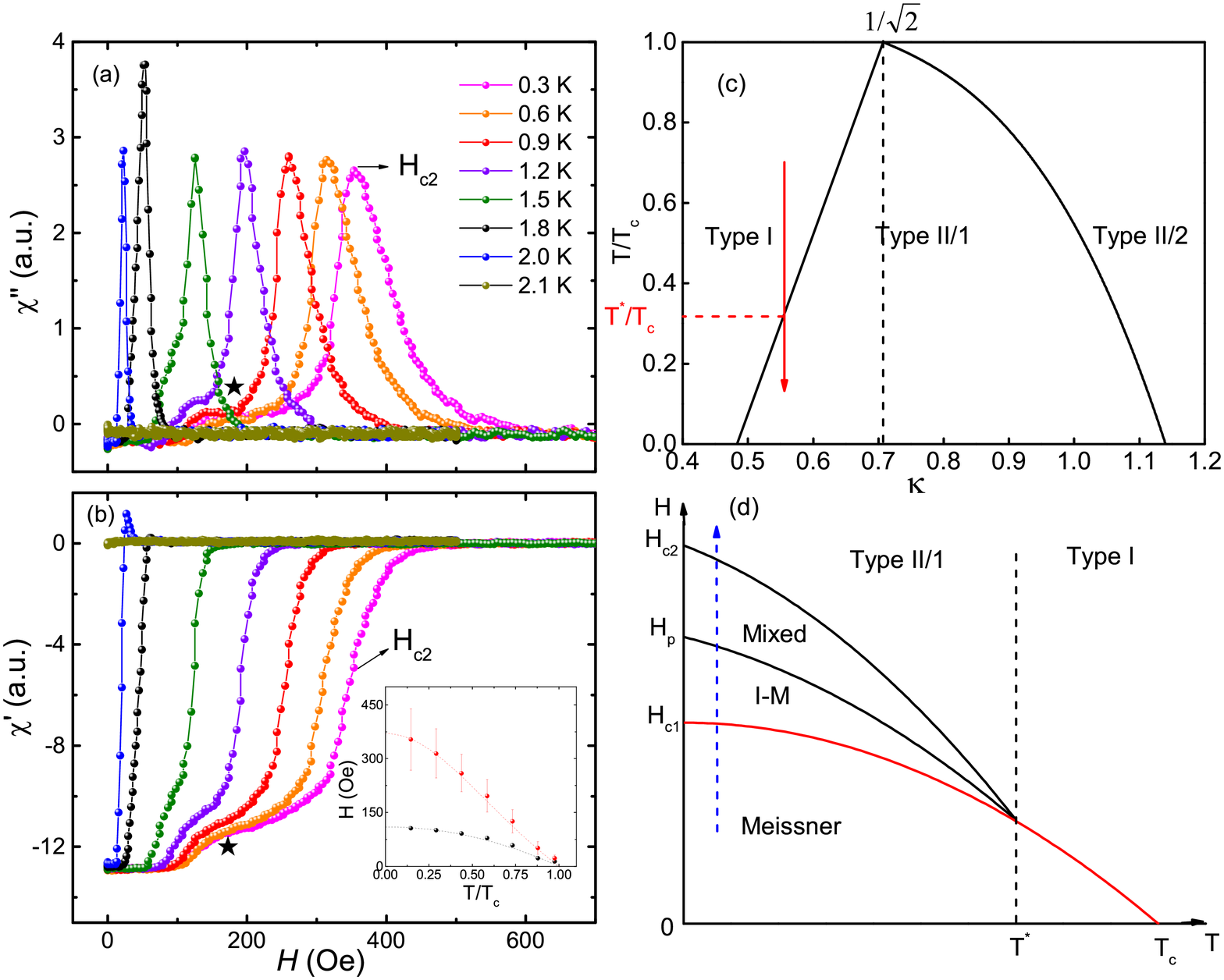}
\vspace{-12pt} \caption{\label{Figure4}(color online) (a) and (b) The field dependent ac susceptibility of NbGe$_2$ at different temperatures with its imaginary and real parts, respectively. Black stars mark the small plateau in ${\chi}$${''}(H)$ and a small step in ${\chi}$${'}(H)$ in the field region of 100 - 200 Oe. The upper critical field at 0.3 K is marked by black arrows at the peak of ${\chi}$${''}(H)$ and in ${\chi}$${'}(H)$. In the inset of (b), the peak of ${\chi}$${''}(H)$ is defined as upper critical field $H_{c2}$ (red dot) with its full width at half maxima (FWHM) taken as the error bar (vertical line).  The lower critical field $H_{c1}$ is defined where ${\chi}$${'}$ starts to increase from the minimum (black dot). $H_{c2}$ and $H_{c1}$ follows different temperature-dependent GL formula (dashed lines). (c) Phase diagram of superconductors with type I, type-II/1 and type-II/2 superconductor regions in the T/T$_c$-$\kappa$ plane. The red line shows a transition from type-I to type-II/1 superconductor at $T^{*}$ with decreased temperatures. (d) H-T phase diagram for $\kappa$ in the range of $\sim0.5$ $<$ $\kappa$ $<$ $1/{\sqrt{2}}$, where it transforms from type-I to type-II/1 superconductor at $T^{*}$. Solid lines show the phase boundaries between Meissner, intermediate-mixed, mixed and normal states, where the first-order phase transition is in red color and second-order phase transitions in black color. (c) and (d) are adapted from Ref. [\onlinecite{JPSJ.85.024715}].}
\vspace{-12pt}
\end{figure}

We have measured the field-dependent ac susceptibility of NbGe$_2$ at different temperatures with its imaginary part ${\chi}$${''}(H)$ and real  part ${\chi}$${'}(H)$ shown in Fig. \ref{Figure4}(a) and (b), respectively, where ${\chi}$${'}$ indicates the shielding ability of superconductor and ${\chi}$${''}$ reflects the magnetic irreversibility \cite{0953-2048/10/8/001}. A differential paramagnetic effect is observed at 2 K, where the real part ${\chi}$${'}$ shows a noticeable positive spike \cite{PhysRev.123.407}. It should originate from the rapid transition from the M = - H to M = 0 state, which is common in type-I and type-II/1 superconductors \cite{PhysRevB.34.4566, JPSJ.85.024715, PhysRevB.85.214526, PhysRevB.96.220506, PhysRevB.99.144519}. With reduced temperatures, the transition range of ${\chi}$${'}$ from the minimum (Meissner state) to zero (normal state) becomes larger, while the peak in ${\chi}$${''}$(H) shifts to higher field and gets broader, showing an increased critical field. The absence of the positive spike in the real part ${\chi}$${'}$ at low temperatures is obviously different from the type-I superconductor behavior \cite{PhysRevB.85.214526, PhysRevB.72.180504}(as demonstrated for both Al and Sn in FigS4 of the Supplemental Material \cite{SI}). At the lowest temperature 0.3 K, the ${\chi}$${'}$(H) keeps constant below 110 Oe with a shielding effect in the Meissner state and continuously increases above 110 Oe till 350 Oe, supporting a type-II superconductivity in NbGe$_2$ at 0.3 K. If we define the peak position in ${\chi}$${''}(H)$ as the upper critical field $H_{c2}$ and the deviation field from the Meissner state as the $H_{c1}$ at 0.3 K, they would be 353 and 106 Oe, respectively, consistent with the values from point-contact spectroscopy. The temperature dependent $H_{c1}(T)$ and $H_{c2}(T)$ are extracted in the same manner as 0.3 K and shown in the inset of Fig. \ref{Figure4}(b), which follow the GL formula $H_{c1}(T)$ = $H_{c1}(0)$$[1-(\frac{T}{T_{c}})^{2}]$ and $H_{c2}(T)$ = $H_{c2}(0)$$\frac{[1-(\frac{T}{T_{c}})^{2}]}{[1+(\frac{T}{T_{c}})^{2}]}$, respectively.

Our MPCS and ac susceptibility measurements have established NbGe$_2$ as a type-I SC at high temperature and type-II SC at 0.3 K. Meanwhile, a transition from type-I to type-II superconductor with decreased temperature has also been confirmed by the specific heat and isothermal magnetization measurements on NbGe$_2$ \cite{PhysRevB.102.235144, PhysRevB.102.064507}. We speculate that NbGe$_2$ might be an intrinsic type-II/1 superconductor as in Fig. \ref{Figure4}(c). As stated earlier, besides the type-I and conventional type-II superconductors classified by the Ginzburg-Landau (GL) parameter $\kappa$, there exists a new branch of type-II/1 superconductor distinct from the conventional type-II superconductor (type-II/2 superconductor), when its $\kappa$ is in the range of $\sim0.5$ $<$ $\kappa$ $<$ $1/{\sqrt{2}}$ . In such a case, a transition from type-I to type-II/1 can occur for the superconductor at $T^{*}$ with decreased temperatures as illustrated in Fig. \ref{Figure4}(c) \cite{PhysRevB.7.136}. Moreover, the superconductor below $T^{*}$ would go through several phases in field, such as Meissner state, intermediate-mixed state, and mixed state before entering into the normal state as in Fig. \ref{Figure4}(d). We note that the $\kappa$ value estimated in Ref. \onlinecite{PhysRevB.102.064507} is only 0.12, much smaller than the phase boundary of 0.5 or $1/{\sqrt{2}}$. On the other hand, the Maki parameter ${\kappa}_{1}$ is defined as ${\kappa}_{1}(T)$ = $\frac{1}{\sqrt{2}}$$\frac{H_{c2}(T)}{H_{c}(T)}$, and ${\kappa}_{1}$ = $\kappa$ with the limit T ${\rightarrow}$ $T_{c}$. In general, the condition ${\kappa}_{1}(T^{*})$ = $\frac{1}{\sqrt{2}}$ determines the transition temperature $T^{*}$ from type-I to type-II SC \cite{PhysRevB.7.136}. We can infer that the $T^{*}$=1.4 K where ${H_{c}(T)}$ deviates from ${H_{c2}(T)}$ \cite{PhysRevB.72.024548}, and ${\kappa}_{1}(0 K)$ ${\simeq}$ 1.14 with the critical field $H_{c2}(0)$ = 360 Oe and $H_{c}(0)$ = 223 Oe from the H-T phase diagram in Ref. \onlinecite{PhysRevB.102.064507}. If we assume a linear temperature dependence of $\kappa_1$(T), the obtained GL parameter $\kappa$ $\equiv$ $\kappa_1(T_c)$ = 0.52 for NbGe$_2$ is right in the region of $\sim0.5$ $<$ $\kappa$ $<$ $1/{\sqrt{2}}$, consistent with its type-II/1 superconducting behaviors. Above the low critical field $H_{c1} \sim$ 150 Oe, the ac susceptibility shows a small step in ${\chi}$${'}(H)$ and a plateau in ${\chi}$${''}(H)$ in Fig. \ref{Figure4}(b), which might be a signal of an abrupt entry of magnetic flux with a discontinuous density in type-II/1 superconductor. However, more microscopic studies are required to further confirm the existence of an intermediate-mixed state and the type-II/1 nature of NbGe$_2$.

In conclusion, we have observed a single full s-wave gap $\Delta$ for the chiral crystal candidate NbGe$_2$ from point-contact spectroscopy and the gap $\Delta$ follows a typical BCS temperature behavior, yielding $\Delta_0 \sim$ 0.32 meV with 2$\Delta_0$/$k\rm_{B}$$T\rm_{c}$ = 3.62 in the weak coupling limit. The differential conductance curves for NbGe$_2$ in magnetic field show a switch from the Meissner state below 150 Oe to a mixed state between 150 and 350 Oe at 0.3 K, characteristic of a possible type-II/1 superconductor at low temperatures.

We are grateful for valuable discussions with Q.H. Chen, C. Cao and Y. Liu. Work at Zhejiang University was supported by the National Key Research \& Development Program of China (Grant No. 2016YFA0300402 and 2017YFA0303101) and the National Natural Science Foundation of China (Grant No. 11674279, No. 11774305, No. 11604291, No. 11374257 and No. 11974306). X.L. would like to acknowledge support from the Zhejiang Provincial Natural Science Foundation of China (LR18A04001).

%\bibliography{NbGe2Ref}% Produces the bibliography via BibTeX.
%

\end{document}